\documentclass[
    ,final            % use final for the camera ready runs
  ,numberedheadings % uncomment this option for numbered sections
  ]
  {aipproc}

\layoutstyle{6x9}

\begin{document}

\title{The broad away side of azimuthal correlations: \\
3 vs 2 final state particles \\
in high energy nuclear collisions}


\author{Alejandro Ayala}{
  address={Instituto de Ciencias Nucleares, Universidad
Nacional Aut\'onoma de M\'exico, \\
Apartado Postal 70-543, M\'exico Distrito Federal 04510, M\'exico.}
}

\author{Jamal Jalilian-Marian}{
  address={Department of Natural Sciences, Baruch College, New York, NY 10010, USA},altaddress={CUNY Graduate Center, 365 Fifth Ave., New York, NY 10016, USA.}
}

\author{Javier Magnin}{
  address={Centro Brasileiro de Pesquisas F\'isicas, CBPF, \\
Rua Dr. Xavier Sigaud 150, 22290-180, Rio de Janeiro, Brazil.}
}

\author{Antonio Ort\'iz}{
  address={Instituto de Ciencias Nucleares, Universidad
Nacional Aut\'onoma de M\'exico, \\ 
Apartado Postal 70-543, M\'exico Distrito Federal 04510, M\'exico.}
}

\author{Guy Pai\'c}{
  address={Instituto de Ciencias Nucleares, Universidad
Nacional Aut\'onoma de M\'exico, \\ Apartado Postal 70-543, M\'exico
Distrito Federal 04510, M\'exico.}
}

\author{Maria Elena Tejeda-Yeomans}{
  address={Departamento de F\'isica, Universidad de Sonora, \\
  Blvd. Luis Encinas y Rosales, Col. Centro, Hermosillo, Sonora 83000, M\'exico.}
}

\begin{abstract}
In high energy heavy ion collisions at RHIC there are 
important aspects of the medium induced dynamics, that are still not well understood. In particular, there is a broadening and even a double hump structure of the away-side peak appearing in azimuthal correlation studies in Au+Au collisions which is absent in p+p collisions at the same energies. 
These features are already present but suppressed in p+p collisions: 2 to 3 parton processes produce such structures but are suppressed with respect to 2 to 2 processes. We argue that in A+A collisions the different geometry for the trajectories of 3 as opposed to 2 particles in the final state, together with the medium induced energy loss effects on the different cross sections, create a scenario that enhances processes with 3 particles in the final state,
which gives on average this double hump structure.
\end{abstract}

\maketitle

%%%%%%%%%%%%%%%%%%%%%%%%%%%%%%%%%%%%%%%%%%%%
%% MAINMATTER
%%%%%%%%%%%%%%%%%%%%%%%%%%%%%%%%%%%%%%%%%%%%

\section{Introduction}

During the last few years there has been many interesting phenomena observed in experiments at RHIC.
The vast majority of such phenomena has been studied and interpreted using well known models that
incorporate energy loss dynamics. The broadening of the away-side peak appearing in azimuthal correlation studies in Au+Au collisions and absent in p+p collisions at the same energies, is one of these remarkable observations. This has yet 
to have a complete understanding that incorporates the medium characteristics and, at the same time, that is coherent
with previous correlation studies. In fact, this particular double hump structure has been the subject of 
different theoretical analysis which are based on the assumption that unlike p+p collisions, 
A+A collisions are strongly influenced by collective phenomena. The purpose of this work is to put 
forward our particular approach to explore the origins of such structure in the away side.
The details of this work have been provided in ~\cite{us1} and soon will be reported in greater extent elsewhere ~\cite{us2}.

\subsection{High $p_T$ partons as probes of the medium}

In relativistic heavy ion collisions, high $p_T$ partons are produced through hard processes 
in the initial binary nucleon collisions and they are the perfect tool to probe the
dynamics of such ephimeral conditions. In fact, just by studying the way partons
hadronize under such conditions, we can obtain information on the nature of the medium. Just after the ion collision, many partons are produced due to the binary collisions between
nucleons. These partons have to travel through the hot and dense medium, before they can
hadronize. If these partons were to hadronize with little 
interaction, then the number of produced high $p_T$ hadrons detected, should scale 
with the number of binary collisions.

\begin{figure}
\includegraphics[height=0.3\textheight]{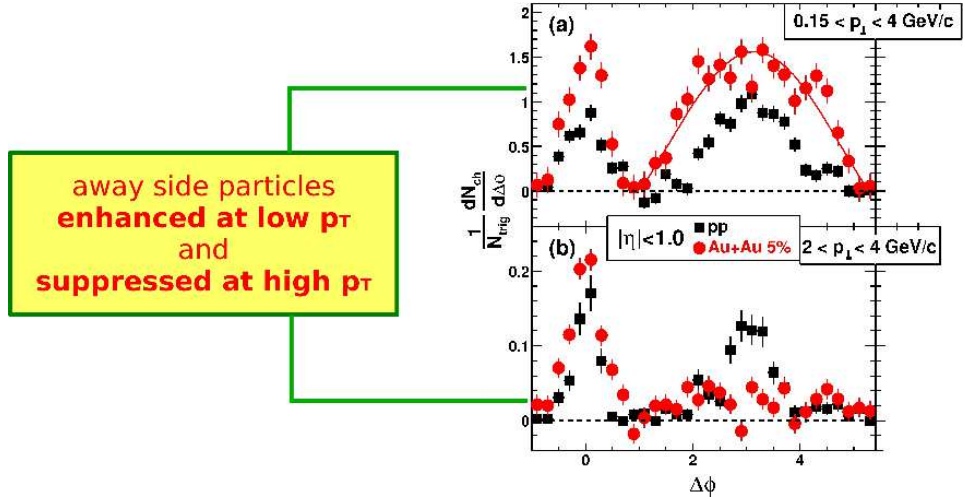}
  \caption{Azimuthal correlations between particle pairs for two $p_T$ ranges in p+p and Au+Au collisions. Reprinted figure with permission from STAR Collaboration (J. Adams et al.), Phys.Rev.Lett. {\bf 95}, 152301, 2005. Copyright 2005 by the American Physical Society.~\cite{STAR2prime}}
  \label{notthere}
\end{figure}

It was quickly realised some time ago that the experimental evidence was telling
a different story: the number of produced high $p_T$ hadrons is reduced 
significantly in these sort of collisions: up to 5 times in most central 
Au + Au collisions ~\cite{STAR1}. This strengthens the idea that the medium
produced in such collisions is opaque for high $p_T$ partons. To study these ideas further, more differential studies were devised
by measuring azimuthal correlations between particle pairs at high $p_T$.
And, as it is shown in FIG.~\ref{notthere}, the near-side correlation is similar 
for the p + p and Au + Au collisions, while the away-side correlation is not 
there for central Au + Au events~\cite{STAR2}.

Many ideas were put forward to encompass a coherent explanation
to what was being observed in these correlation studies. Among others,
there were many \textit{elliptic flow studies} implemented. These studies
are based on the fact that an anisotropy in the momentum distribution of particles
may arise after the initial binary collisions. They consider that the initial 
geometry of the collision region is anisotropic in the azimuthal direction so that  after the interacting system reaches local thermal equilibrium, preassure gradients 
are steeper in the impact parameter direction and these generate the elliptic flow.

So up until then, the elliptic flow together with the two particle 
correlation studies applied on different analisis, gave an
indication that an opaque, strongly interacting partonic matter 
had been created in the high energy Au + Au collisions at RHIC.

\subsection{A puzzle in correlation studies}

In 2005 there was a rich correlation structure in Au + Au vs p + p
reported by the leading collaborations at RHIC. 
They observed a \textit{ridge} and \textit{broad away side} in the
correlation studies performed at the time. More precisely, they
reported an excess yield of correlated particles at $\Delta\phi=0^{\circ}$
and $\Delta\phi\approx 180^{\circ}$ extending out to $\Delta\eta > 2$
~\cite{PHENIX1etal}

\begin{figure}
\includegraphics[height=0.35\textheight]{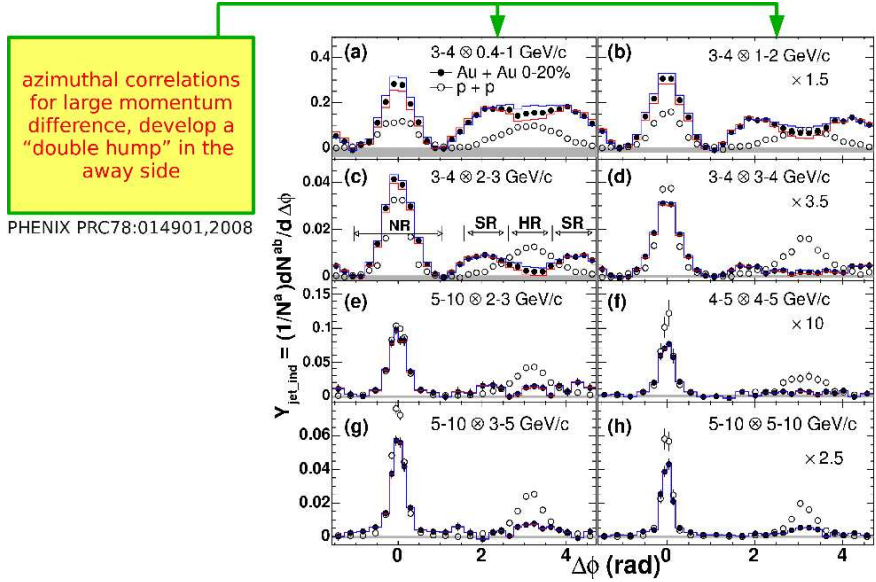}
  \caption{Two particle correlations for p+p and Au+Au in different
  $p_T$ bins. Reprinted figure with permission from PHENIX Collaboration (A. Adare et al.), Phys. Rev. C {\bf 78}, 014901, 2008. Copyright 2008 by the American Physical Society. ~\cite{PHENIX1}}
  \label{headshoulder}
\end{figure}

\begin{figure}
\includegraphics[height=0.3\textheight]{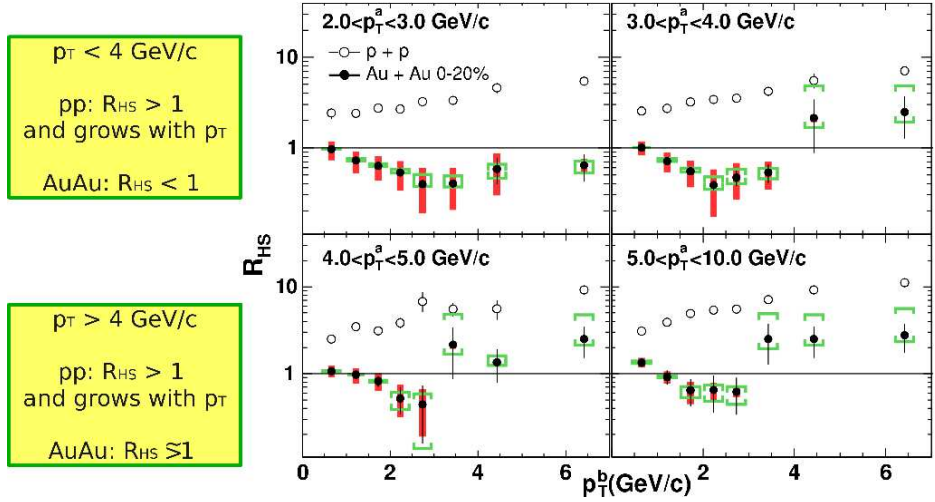}
  \caption{Ratio $R_{HS}$ (\textit{head/shoulder}) for different
  $p_T$ bins. Reprinted figure with permission from PHENIX Collaboration (A. Adare et al.), Phys. Rev. C {\bf 78}, 014901, 2008. Copyright 2008 by the American Physical Society. ~\cite{PHENIX1}}
  \label{ratioRHS}
\end{figure}

In FIG.~\ref{headshoulder} we can see that, to analize the origins 
of such structure the PHENIX collaboration identified 
\textit{head} and \textit{shoulders}
regions and defined a ratio $R_{HS}$ (\textit{head/shoulder}).
Looking closely at the ratio $R_{HS}$ in FIG.~\ref{ratioRHS}, we can see
that for p+p collisions $R_{HS}$ grows with $p_T$, which can be
interpreted as a production of a narrower jet, whereas for Au + Au, jet 
fragmentation dominates at high $p_T$ over medium effects.

Several theoretical models have been proposed to look for the origins of 
such structures, among others the literature focuses mainly on

\begin{itemize}

\item Mach or Cerenkov cone due to medium reaction~\cite{mach}: 
the \textit{double hump} can be explained by considering how the medium 
reacts to the passing of a fast parton. In principle this would produce
such structures since the medium would eject two bunches of hadrons in the away side. 

\item Triangularity and triangular flow~\cite{tflow}: 
the \textit{double hump} can be explained by considering
event-by-event fluctuations in the initial collision geometry as a next 
order collective flow effect, after considering the elliptic flow effects.

\end{itemize}

We can see that all of the theoretical models rely on the description
of such away-side structures as the manifestation of emergent behaviour
due to the collective (e.g. triangular flow: relies crucially on the 
existence of initial geometry fluctuations). In fact, there are 
a couple of recent reviews on the models posed to solve this puzzle: 
up until 2009, J. L. Nagle ~\cite{Nagle} argued that "...none of the 
theoretical models are succesful to describe
all the special characteristics of these structures..."
and this year, M. J. Tannenbaum ~\cite{paradigm} argued that "...no clear 
paradigm has emerged for the two-lobed wide away-jet structure..."

\section{Solving the puzzle: our proposal}

In a nutshell, our proposal~\cite{us1,us2} regarding the origins of the 
double-hump in the away side of two-particle correlation studies in A + A collisions, is as follows

\begin{itemize}
\item[$\surd$] we account for the medium induced energy loss effects on the calculation of $2\to \{2, ~3\}$ cross sections

\item[$\surd$] we also take into account the different path lengths for the trajectories of 3 as opposed to 2 particles in the final state of an A + A collision

\item[$\surd$] finally, considering that one particle is absorbed by the medium and the other one punches through

\item[$\rightarrow$] this gives on average, a double hump structure
\end{itemize}

\begin{figure}
\label{dist1}
\includegraphics[scale=0.3]{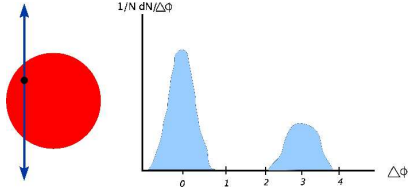}
  \caption{Schematic plot of a correlation study for a $2\to 2$ process.}
\end{figure}

\begin{figure}
\label{dist2}
\begin{tabular}{c}
\includegraphics[scale=0.3]{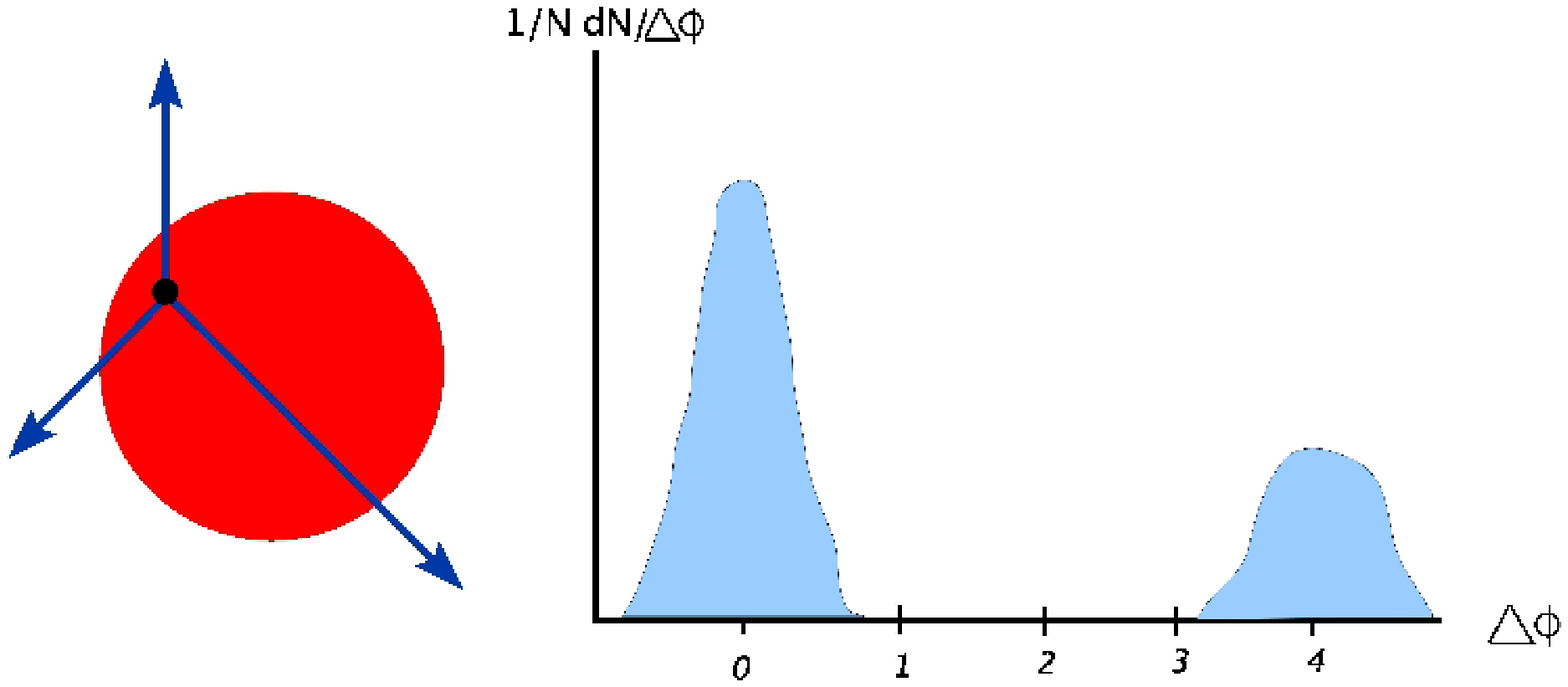} \hspace{1cm} \includegraphics[scale=0.3]{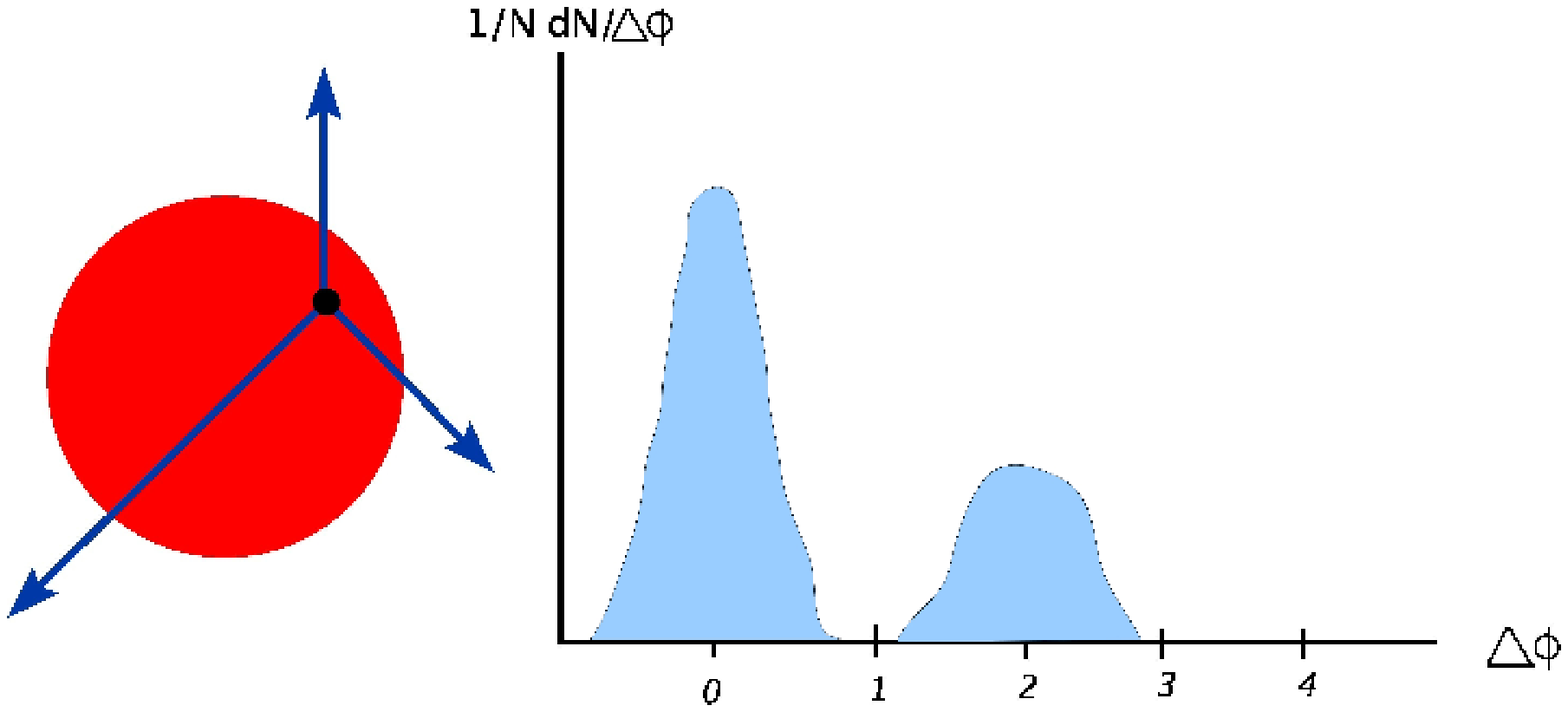} \\
\includegraphics[scale=0.3]{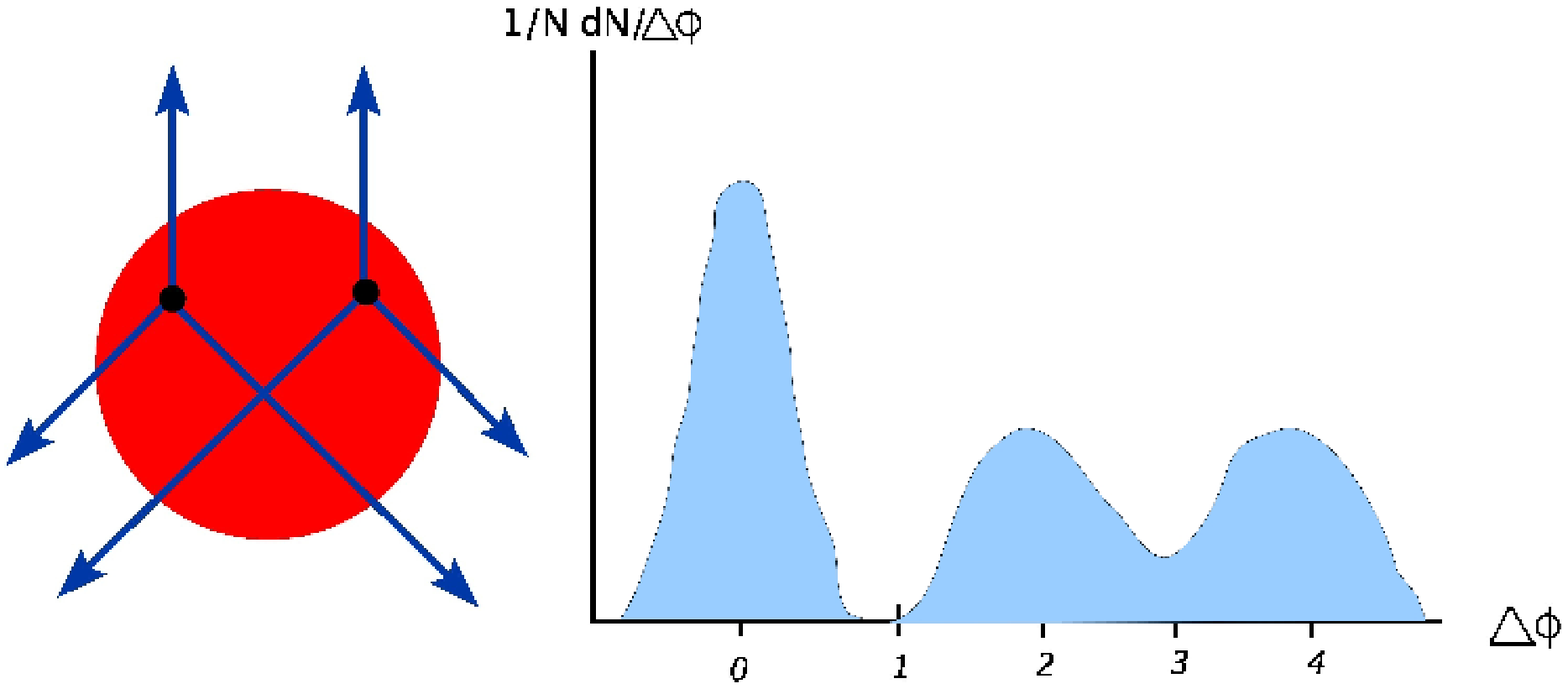} 
\end{tabular}

\caption{From left to right: schematic plots of a correlation studies for a $2\to 3$ process, with particle at angle $2\pi/3$ absorbed, with particle at angle $4\pi/3$ absorbed and combined.}
\end{figure}

In other words, when we have a $2 \to 2$ scattering process, we have a correlation
between the leading and the away side that looks, schematically as shown in FIG.~\ref{dist1}. There, one expects the defined humps to be around $0$ and $\pi$. 
But when one considers the possibility of a significant contribution 
to this correlation studies, coming from a promotion of the $2 \to 3$ scattering
processes, on average one has a double hump structure (roughly at around $2\pi/3$ and $4\pi/3$), as shown in FIG.~\ref{dist2},
again schematically.

As we summarized before, in order to build our proposal we need
to account for energy loss efects in the calculation of
the cross sections that are relevant to this study. In particular
the $2\to 2$ and $2\to 3$ differential cross sections can be
represented as follows:
\begin{eqnarray}
   \frac{d\sigma^{pp \rightarrow h1\, h2\, X}}{d\, y_1 d\, y_2 d\,
   h_{1t}  d\, h_{2t} d\, \phi_2 } & \propto & \int d z_2  |{\mathcal{M}^{2\to 2}}|^2 
   f_{i/p} (x_1, \mu^2 )  f_{j/p} (x_2, \mu^2 ) \nonumber \\
   && \times D_{h1/k}^0 (z_1, \mu^2 ) D_{h2/m}^0 (z_2, \mu^2 ) \nonumber  \\ ~ \\
   \frac{d\sigma^{pp \rightarrow h1\, h2\, h3\, X}}{d\, y_1 d\, y_2 d\, y_3 d\,
   h_{1t}  d\, h_{2t} d\, h_{3t} d\, \phi_2 d\, \phi_3 } &\propto & 
   \int d z_3  |{\mathcal{M}^{2\to 3}}|^2 f_{i/p} (x_1, \mu^2 )  f_{j/p} (x_2, \mu^2 ) \nonumber \\
   && \times D_{h1/k}^0 (z_1, \mu^2 ) D_{h2/m}^0 (z_2, \mu^2 ) D_{h3/n}^0 (z_3, \mu^2 ).
   \nonumber \\
\end{eqnarray}
We can see that for both p + p and Au + Au collisions we need to calculate the scattering amplitudes ${\mathcal{M}^{2\to 2}}$, ${\mathcal{M}^{2\to 3}}$ and use them
together with their corresponding parton distribution functions $f_{i/p}$ and 
parton fragmentation functions $D_{h/k}$, when integrating over the appropiate phase space.

\begin{figure}
\label{amplitudes}
\begin{tabular}{c}
\includegraphics[scale=0.6]{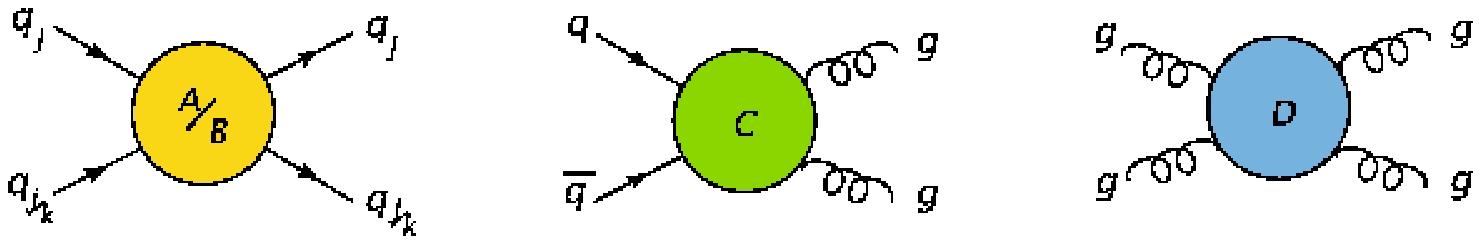} \\
\includegraphics[scale=0.5]{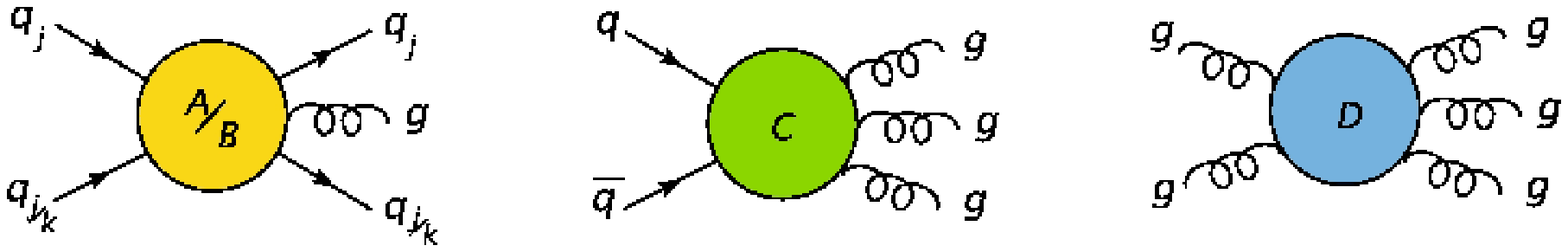} 
\end{tabular}
  \caption{Scattering amplitudes for $2 \to 2$ and $2 \to 3$ processes considering all possible external states with crossings.}
\end{figure}

For the lowest order amplitudes ${\mathcal{M}^{2\to 2}}$ and ${\mathcal{M}^{2\to 3}}$
we need to consider 4 classes of diagrams and their crossings at the parton level
(see for example~\cite{ellis}), as is schematically shown in FIG.~\ref{amplitudes}.

\begin{figure}
\label{ppxsec}
\includegraphics[scale=0.5]{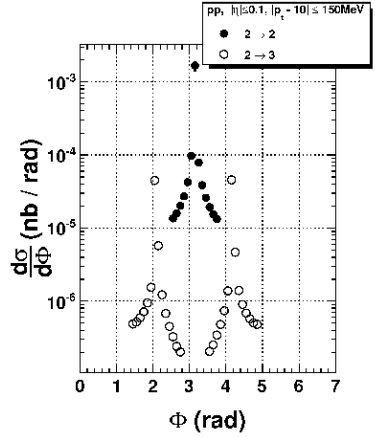}
  \caption{$2, 3$ final state particle differential cross section at $\sqrt{S} = 200$ GeV for p + p collisions~\cite{us1}}
\end{figure}

In our calculation, we use the CTEQ6 parametrization for the parton 
distribution functions~\cite{cteq6} and the KKP parametrization~\cite{kkp}
for the (unmodified) parton fragmentation funtions together with
LO-DIPHOX~\cite{diphox} to compare with the $2 \to 2$ result.
In the case of p + p collisions the $2, 3$ final state particle
differential cross section at $\sqrt{S} = 200$ GeV is shown in FIG.~\ref{ppxsec}
for the away side hadrons as a function of the azimuthal angle.
We focus on midrapidity region ($y_i=0$) and as an example that simplifies the calculation, consider a situation in which all hadrons carry $10$ GeV/c. 
We can see
that we have two well defined peaks at $\Delta \phi = \pi$ 
($\Delta \phi = 2\pi/3, 4\pi/3$), for a $2$ ($3$) particle final state.

In order to describe the same observables but for a Au + Au collision, we now use modified parton fragmentation functions proposed by Zhang et al~\cite{modkkp}, 
to account for the effects of the medium
on the propagation of the produced partons. These fragmentation functions are
parametrized as follows:
\begin{eqnarray}
   D_{h/i}(z_i,\mu^2)& = & 
   (1-e^{-{ \langle \frac{L}{\lambda}\rangle}}) 
   \left[ \frac{ z_i^\prime}{z_i}D^0_{h/i}({ z_i^\prime},\mu^2) +
   { \langle \frac{L}{\lambda}\rangle} \frac{ z_g^\prime}{z_i} 
   D^0_{h/g}({ z_g^\prime},\mu^2)\right]
   + e^{-{ \langle\frac{L}{\lambda}\rangle}} D^0_{h/i}(z_i,\mu^2) \nonumber \\
\end{eqnarray}
where
\begin{itemize}
\item[] { $z_i^\prime= \frac{h_t}{(b_{ti}-\Delta E_i)}$} is the rescaled momentum
fraction of the leading parton with flavor $i$,
\item[] { $z_g^\prime=\langle \frac{L}{\lambda} \rangle \frac{b_t}{\Delta E_i}$} 
is the rescaled momentum fraction of the radiated gluon,
\item[] { $\langle \frac{L}{\lambda}\rangle$} is the average number of scatterings
\item[] and the average radiative parton energy loss is taken to be 
$$\Delta E \propto \langle \frac{dE}{dL} \rangle_{1d} \, \int_{\tau_0}^{\infty} d
   \tau \Delta \tau \, 
   \rho_g (\tau, \vec{r}_t + \vec{n} \tau). $$
\end{itemize}   

\begin{figure}
\label{AAxsec}
\includegraphics[scale=0.5]{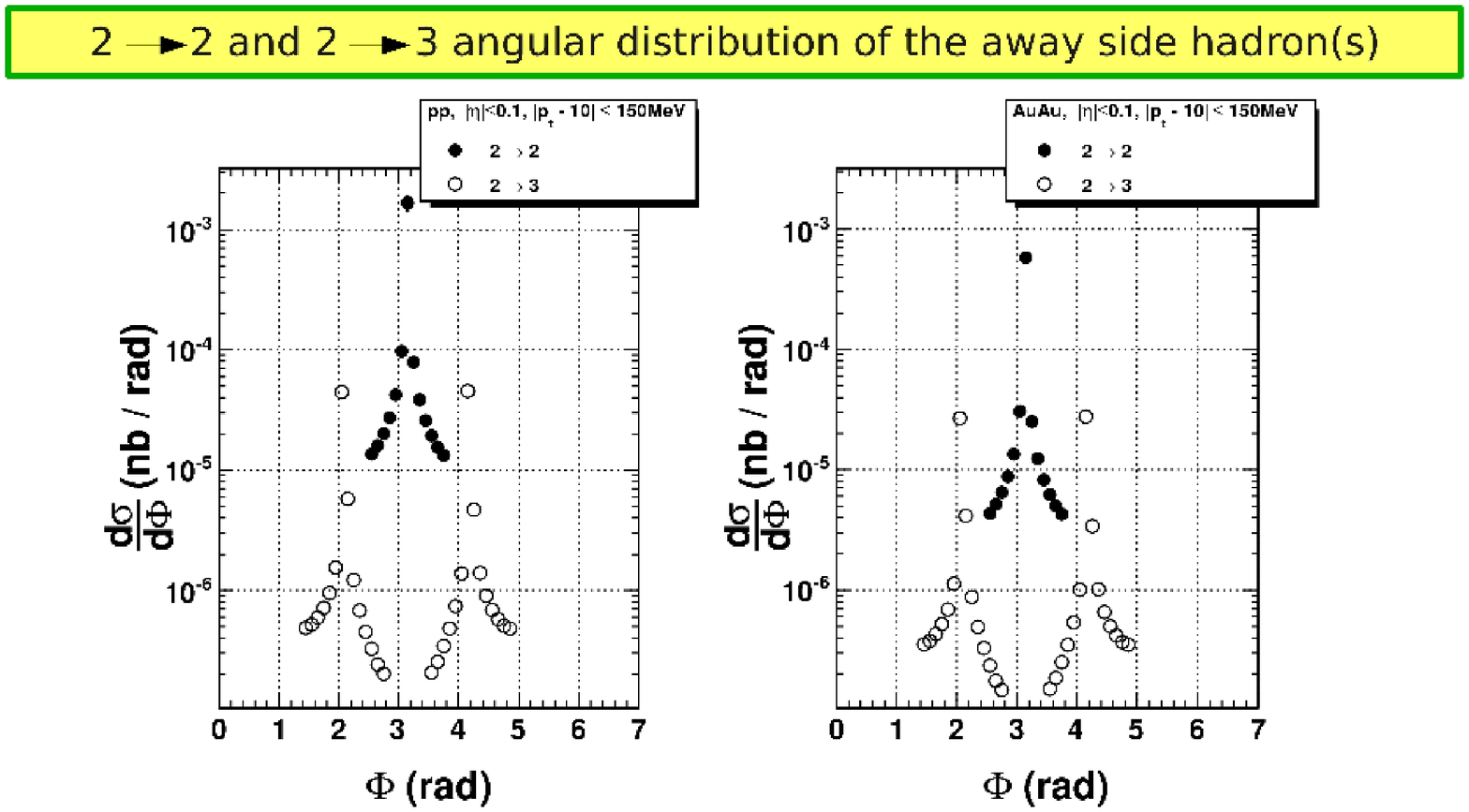}
\caption{$2, 3$ final state particle differential cross section at $\sqrt{S} = 200$ GeV for Au + Au collisions~\cite{us1}}
\end{figure}

Also $\vec{r}_t $ is the transverse plane location of the hard scattering where the
partons are produced, $\vec{n}$ is the direction in which the produced hard parton travels in the medium and for most central collisions $\vec{b}_{\perp} = 0$.
In our calculation, we use $\langle \frac{L}{\lambda}\rangle$, $\langle \frac{d E}{d L}\rangle_{1d}$ and $\rho_g$ (motivated by the geometry) as suggested in ~\cite{modkkp}. Taking into account the modified parton fragmentation function, in FIG.~\ref{AAxsec} 
we show the appropiate differential cross section for a $2, 3$ particle final state 
in Au + Au collisions, as was done previously for p + p. Notice that in both cases
the 3 hadron production cross section is suppressed with respect to the 2 hadron final state result. However, this suppression is smaller in A + A
collisions than in p + p collisions. Dividing this ratio in A + A to that in
p + p, we get as a function of $\phi$ approximately a constant:  
 \begin{eqnarray}
 \frac{\mbox{Au + Au}:~ \frac{2\to 3}{2\to 2}}{\mbox{p + p}:~\frac{2\to 3}{2\to 2}}~  \sim  2.26.
\end{eqnarray}

\begin{figure}
\label{nuclearover}
\includegraphics[scale=0.5]{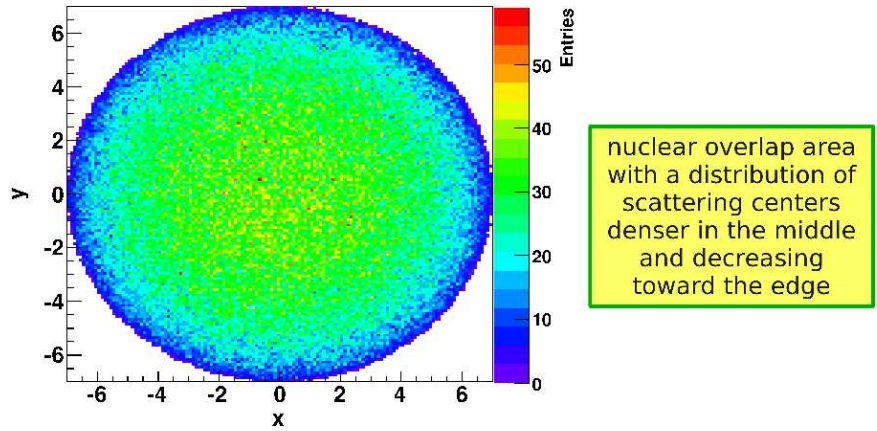}
  \caption{Distribution of scattering centers in the nuclear overlap area~\cite{us1}}
\end{figure}

Notice that the sole ingredient that induced this enhancement from the 
calculation of $2\to 3$ vs $2 \to 2$ in Au + Au to that of p + p collisions 
is the energy loss of partons that hadronize collinearly. So this must be 
correlated to the different geometry for the trajectories of 3 as opposed 
to 2 particles in the final state. In order to test this idea we computed 
the distribution of path lengths with two
and three hadrons in the final state by taking a nuclear overlap area
with a distribution of scattering centers denser in the middle and decreasing
toward the edge, as shown in FIG.~\ref{nuclearover}. 

\begin{figure}
\label{paths}
\includegraphics[scale=0.5]{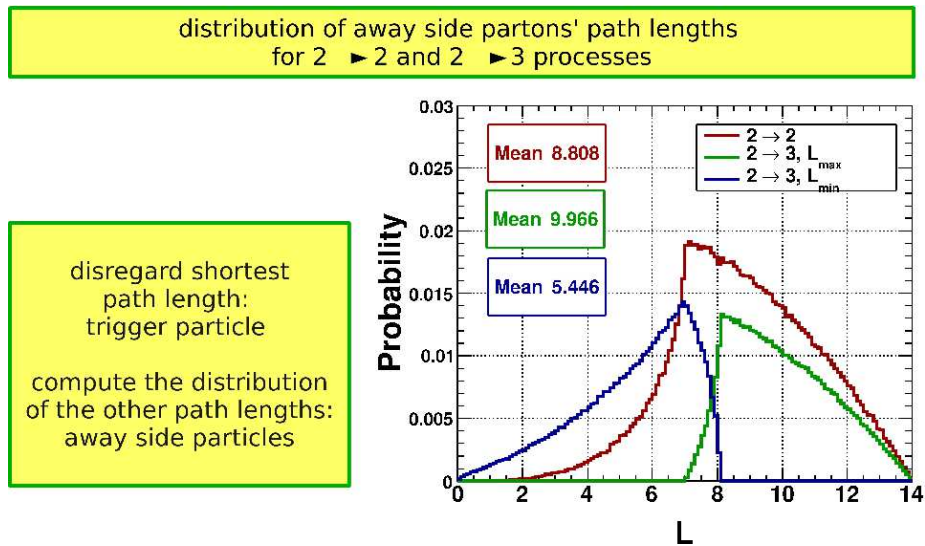}
  \caption{Distribution of path lenghts for the away side particles~\cite{us1}}
\end{figure}

In each case we disregard the path length 
that would correspond to the trigger particle and 
in FIG.~\ref{paths} we compute the distribution of
the path lengths corresponding to the away side
particles. As you can see, when there are three hadrons in the final state,
the large (short) path length in the away side is greater (smaller) than the
case of the away side particle when there are two hadrons in the final state,
 i.e. ${\mbox{L}_{\mbox{min}}^{2 \to 3}} < { \mbox{L}^{2 \to 2}} < { \mbox{L}_{\mbox{max}}^{2 \to 3}}$. This means that for the case of 
three particles in the final state, even if one of the non-leading particles
with the largest path length gets absorbed by the medium, the
remaining particle has a larger probability of punching through than in the
case when one has two particles in the final state. So, in processes with
3 particles in the final state, there is a large probability to have
one of the two away side particles being absorbed and the other randomly
getting out, producing on the average, a double hump structure in the correlation
studies.

\section{Final remarks}

We want to emphasize that $2\to 3$ processes have to be accounted for in current heavy ion experiments, given that the medium levels out
the $2\to 2$ processes rates. In other words, their observation 
should be enhanced with respect to suppressed $2\to 2$ processes.
Moreover, we claim that this effect may have bearing on the away side shape 
for different kinematical cuts in Au + Au collisions. We realize that we need three-particle-correlation measurements to distinguish 
between ours and other scenarios that might be responsible for this shape in 2 particle correlation studies and in fact we are working towards 
providing our predictions in this context.

%%%%%%%%%%%%%%%%%%%%%%%%%%%%%%%%%%%%%%%%%%%%%%%%
%% BACKMATTER
%%%%%%%%%%%%%%%%%%%%%%%%%%%%%%%%%%%%%%%%%%%%%%%%

\begin{theacknowledgments}
The authors thank the organizers of the \textit{5th International Workshop on High $p_T$ physics at the LHC, 2010} at 
ICN-UNAM, Mexico, where this work was presented. M.E.T. appreciates the support provided by \textit{RedFAE} CONACyT and Departamento de F\'isica at Universidad de Sonora. 
\end{theacknowledgments}

\bibliographystyle{aipproc}   % if natbib is available

\end{document}